\documentclass[sigconf]{acmart}

\AtBeginDocument{%
  }

\setcopyright{acmcopyright}
\copyrightyear{2018}
\acmYear{2018}
\acmDOI{XXXXXXX.XXXXXXX}

\acmConference[Conference acronym 'XX]{Make sure to enter the correct
	conference title from your rights confirmation emai}{June 03--05,
	2018}{Woodstock, NY}
\acmPrice{15.00}
\acmISBN{978-1-4503-XXXX-X/18/06}

\usepackage{multirow}
\usepackage{subfigure}
\usepackage{amsfonts}
\usepackage{makecell}
\begin{document}

\title{Auditory Attention Decoding with Task-Related Multi-View Contrastive Learning}

\author{Xiaoyu Chen}
\authornote{Both authors contributed equally to this research.}
\orcid{0000-0001-8945-3150}
\affiliation{%
  \institution{Laboratory of Brain Atlas and Brain-Inspired Intelligence, State Key Laboratory of Multimodal Artificial Intelligence Systems, Institute of Automation, Chinese Academy of Sciences}
  \institution{\& School of Artificial Intelligence, University of Chinese Academy of Science}
  \city{Beijing}
  \country{China}
}
\email{chenxiaoyu2022@ia.ac.cn}

\author{Changde Du}
\authornotemark[1]
\orcid{0000-0002-0084-433X}
\affiliation{%
  \institution{Laboratory of Brain Atlas and Brain-Inspired Intelligence, State Key Laboratory of Multimodal Artificial Intelligence Systems, Institute of Automation, Chinese Academy of Sciences}
  \city{Beijing}
  \country{China}
}
\email{changde.du@ia.ac.cn}

\author{Qiongyi Zhou}
\orcid{0000-0001-9498-9523}
\affiliation{%
  \institution{Laboratory of Brain Atlas and Brain-Inspired Intelligence, State Key Laboratory of Multimodal Artificial Intelligence Systems, Institute of Automation, Chinese Academy of Sciences}
  \institution{\& School of Artificial Intelligence, University of Chinese Academy of Science}
  \city{Beijing}
  \country{China}
}
\email{zhouqiongyi2018@ia.ac.cn}

\author{Huiguang He}
\orcid{0000-0002-0684-1711}
\authornote{corresponding author.}
\affiliation{%
  \institution{Laboratory of Brain Atlas and Brain-Inspired Intelligence, State Key Laboratory of Multimodal Artificial Intelligence Systems, Institute of Automation, Chinese Academy of Sciences}
  \institution{\& School of Artificial Intelligence, University of Chinese Academy of Science}
  \city{Beijing}
  \country{China}
}
\email{huiguang.he@ia.ac.cn}

\renewcommand{\shortauthors}{Xiaoyu Chen, Changde Du, Qiongyi Zhou, \& Huiguang He}

\begin{abstract}
  The human brain can easily focus on one speaker and suppress others in scenarios such as a cocktail party. Recently, researchers found that auditory attention can be decoded from the electroencephalogram (EEG) data. 
  However, most existing deep learning methods are difficult to use prior knowledge of different views (that is attended speech and EEG are task-related views) and extract an unsatisfactory representation. 
  Inspired by Broadbent's filter model, we decode auditory attention in a multi-view paradigm and extract the most relevant and important information utilizing the missing view. Specifically, we propose an auditory attention decoding (AAD) method based on multi-view VAE with task-related multi-view contrastive (TMC) learning. Employing TMC learning in multi-view VAE can utilize the missing view to accumulate prior knowledge of different views into the fusion of representation, and extract the approximate task-related representation.
  We examine our method on two popular AAD datasets, and demonstrate the superiority of our method by comparing it to the state-of-the-art method.
\end{abstract}




\maketitle

\section{Introduction}
\begin{figure}[h]
	\centering
	\includegraphics[width=\linewidth]{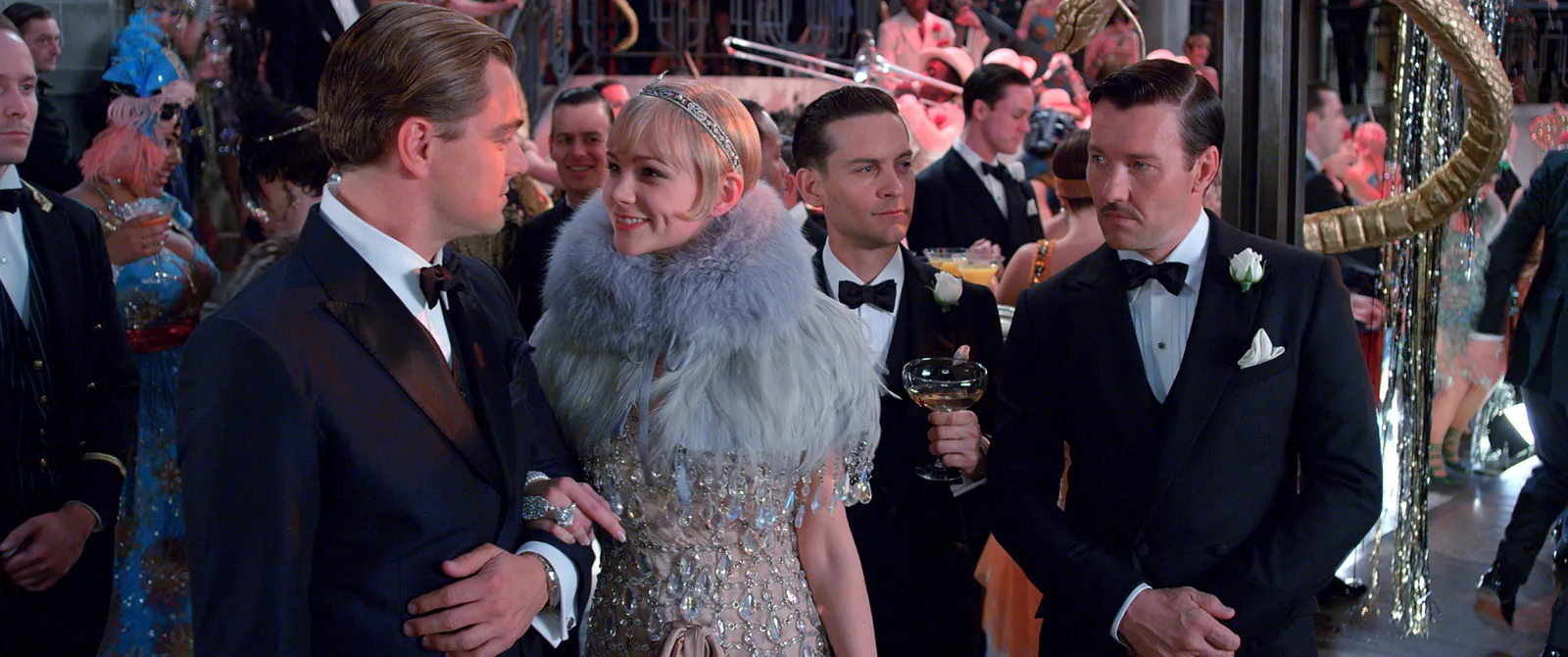}
	\caption{In a cocktail party, one can focus on the interested speaker while ignoring other interference sounds. (Image from The Great Gatsby: Warner Bros. Pictures and Roadshow Films.)}
	\label{fig:cocktail}
\end{figure}

In the acoustic environments people face every day, one’s brain can focus auditory attention on a particular stimulus while filtering out other stimuli. For example, people can focus on their interest speaker during a cocktail party (see Figure \ref{fig:cocktail}).  This marvel phenomenon, the cocktail party effect \cite{cherry1953some},  has attracted long-standing research interest \cite{mcdermott2009cocktail, conway2001cocktail, golumbic2013mechanisms}. And the mechanism behind it is often called selective auditory attention \cite{dai2018neural, mesgarani2012selective, golumbic2013mechanisms}.

Recently, with the development of the brain-computer interface (BCI), researchers are interested in decoding auditory attention through brain activities, which is known as auditory attention decoding (AAD). Auditory attention can be decoded from several brain signals, such as 
electrocorticography (ECoG) \cite{mesgarani2012selective}, magnetoencephalography (MEG) \cite{ding2012neural} or electroencephalography (EEG) \cite{o2015attentional}. Since it is economical and non-invasive, methods based on EEG have the most promising application potential and may affect hearing aids and active noise cancellation (ANC) headphones in the future.

\begin{figure}[h]
    \centering
    \includegraphics[width=\linewidth]{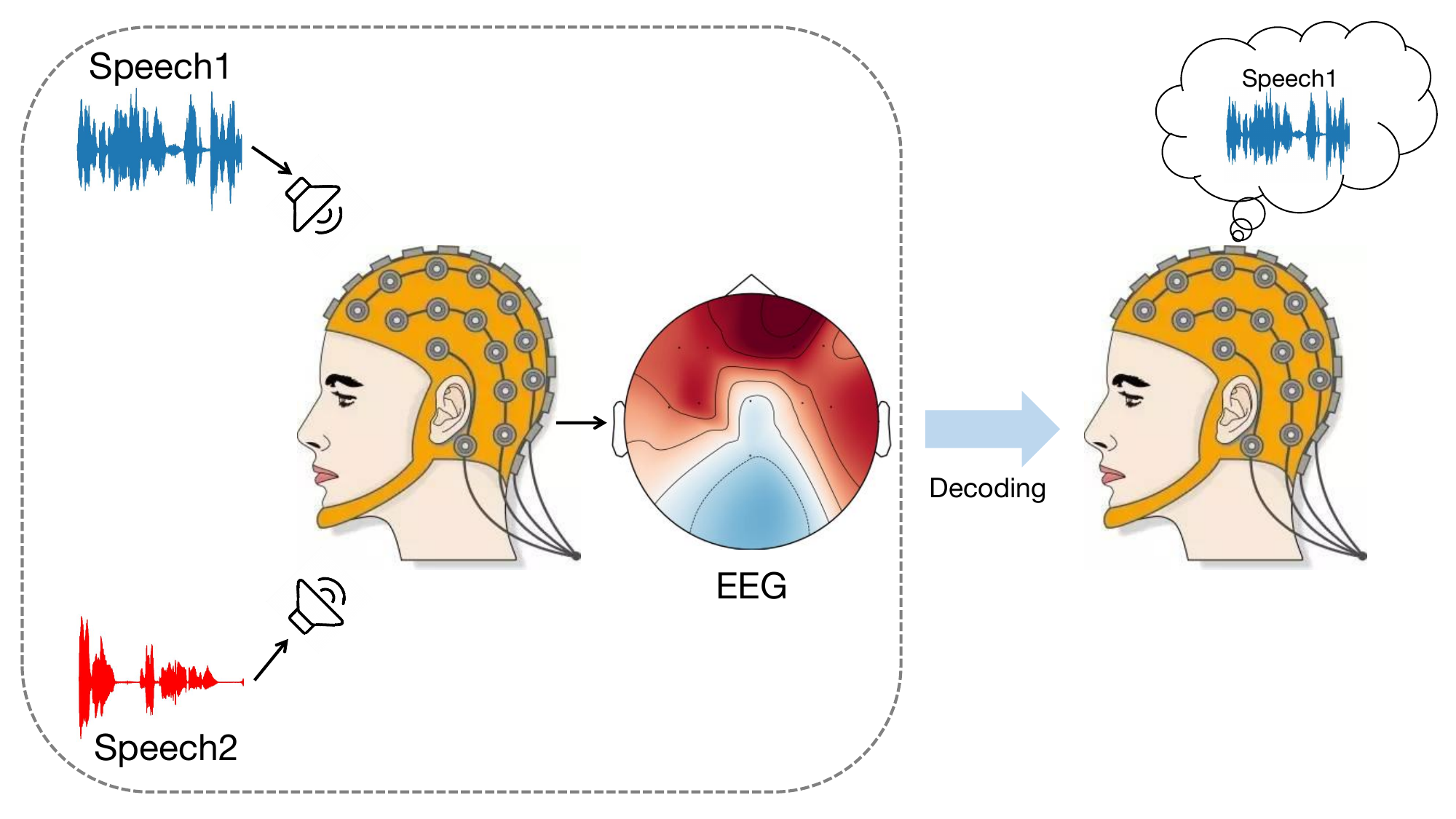}
    \caption{Auditory attention decoding in the dual-speaker scenario. The subject will hear two different speeches when during the EEG recording. The AAD methods can infer the subject's attention based on the speeches and EEG.}
    \label{fig:dual_speaker}
\end{figure}

In the dual-speaker scenario, which is the most popular experimental form in recent AAD research, the subject will hear two different speeches and choose one speech as the attended one actively or passively (see Figure \ref{fig:dual_speaker} for an example). And the task of AAD methods is to infer the subject’s attended speech  based on the EEG and two speeches. Most existing AAD methods resort to extracting the representation using all the information in the data \cite{kuruvila2021extracting,cai2021eeg,xu2022auditory}. However, the prior knowledge of the AAD task is that the attended speech and the EEG are two related views, which contain information about auditory attention. And such a relationship has been ignored in the existing deep learning AAD methods. According to Broadbent's filter model \cite{broadbent1957mechanical, broadbent2013perception}, the attentional processing system in the human brain has an early selection process that uses a selective filter to avoid unrelated information getting involved in the higher-level processing. With this filtering mechanism, our human brain can have a remarkable capability to pay attention to a particular sound source and ignore surrounding noise, such as focusing on the attended speaker at a cocktail party intentionally or making conversation with your friends on a noisy train. Therefore, we argue that the representation should be extracted from the task-related part of the data. 

Inspired by Broadbent's filter model, we developed our method in a multi-view structure and filter the unrelated information when fusion the representation. Specifically, our work refers to the EEG and speeches as different views of data and decoding auditory attention based on multi-view variational autoencoder (VAE) \cite{kingma2013auto,sutter2021generalized,wu2018multimodal,shi2019variational}. The multi-view VAE will transform the different views of data into different single-view representations at the beginning, and fuse them to a common representation space. Then several decoders will be trained, and map the representation from the common space back to different views of data. The common space can encode the distribution of multi-view data effectively after training. Since that, the multi-view VAE can leverage the underlying relationship between different views of data and improve the performance of AAD methods. 

When implementing the multi-view VAE in the AAD task, a critical problem is how to effectively utilized the prior knowledge about different views of data. In fact, the information about selective attention is contained in the attended speech and EEG, which we called task-related representations (or views). So it is important to retain more information from the task-related views and minimize the interference of task-unrelated views (unattended speech) during the fusion of single-view representations. Since the multi-view VAEs support learning a representation of data with the missing view, a straightforward thought is fusing the task-related representation on attended speech and EEG views. But unfortunately, the multi-view VAE needs the fused common representation to carry out the AAD task, while fusing a task-related common representation needs the result of AAD. This dilemma causes a great obstacle in the application of task-related representation in the AAD task. To solve this problem, we propose task-related multi-view contrastive (TMC) learning to extract the approximate task-related representation.

The TMC learning consists of two main ideas: 1) utilizing the support of missing view in multi-view VAE to fuse a task-related representation and 2) approximate task-related representation using contrastive learning. Specifically, we first fuse the task-related representation based on the attended speech and EEG according to the label in the training stage. Since the label is unavailable in testing, we then fuse a complete representation, which depends on all the speeches and EEG, and align the complete representation with the task-related one using contrastive learning. Through that, the TMC can approximate the task-related representation by the complete one. Since the fusion of complete representation does not need label information, we can get an approximate solution to the non-trivial problem above. 

\begin{figure*}[t]
    \centering
    \includegraphics[width=\linewidth]{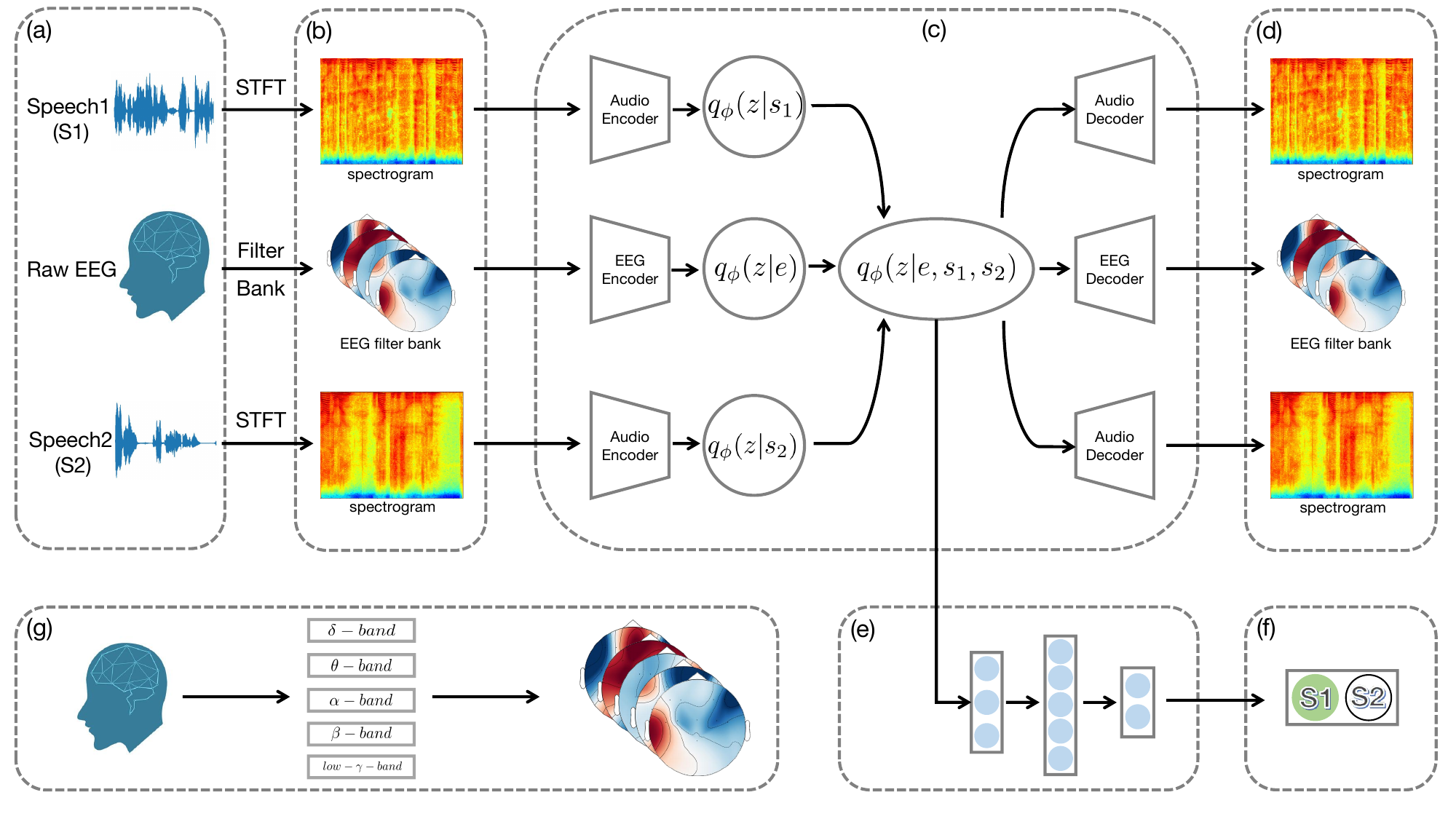}
    \caption{Overview of our model architecture. (a) The raw EEG and speech stimuli. (b) Preprocessed EEG and speech stimuli. We use the spectrogram for speech stimuli and the filter bank feature for the EEG. (c) We use a multi-view VAE architecture to extract the representation from the EEG and speech stimuli. (d) The reconstructed inputs. (e) Our classification network is constructed by a three-layer MLP. (f) The classifier will output a one-hot vector that indicates the AAD result. (g) The process of extracting filter bank EEG feature. The raw EEG will be passed through different band-pass filters to form the filter bank feature.}
    \label{fig:network}
\end{figure*}

{\bf Contributions.} Our main contributions are: 1) By applying the multi-view VAE, we construct our method to exploit the information in the multi-view data and learn a more comprehensive representation (see figure \ref{fig:network} for an overview of the framework). 2) We propose task-related multi-view contrastive (TMC) learning which can utilize the prior knowledge about different views of data to learn an approximate task-related representation effectively. 3) The experiments show that our method is comparable to or much better than the state-of-the-art methods on two popular AAD datasets.

\section{Related Work}


\subsection{Auditory Attention Decoding}
\subsubsection{The Traditional Methods.} The traditional correlation-based methods can be divided into two categories, forward encoding methods and backward decoding methods. The ideas of forward encoding methods and backward decoding methods are consistent with neural encoding and decoding, that is, predicting the brain activities (EEG) given stimuli, or reconstructing the stimuli given brain activities \cite{geirnaert2021electroencephalography}. In the forward encoding methods, an encoder will be trained to infer the EEG given different speech signals \cite{alickovic2019tutorial,wong2018comparison} and correlated to the real EEG to decide the attended speech. While for the backward decoding methods \cite{de2018decoding,das2018eeg,katthi2020deep,de2021auditory} , a stimulus reconstruction pattern is wildly accepted \cite{o2015attentional,mirkovic2015decoding,biesmans2016auditory}. The envelope of attended speech will be reconstructed from the EEG, and then compare with all the speeches using the Pearson correlation coefficient. The traditional methods are mostly based on a linear model, which fails to capture the nonlinear characteristics of the human auditory system \cite{zwicker2013psychoacoustics}. 

\subsubsection{Deep Learning Methods.} With the applications of deep learning in the brain-computer interface, many works decoded auditory attention using deep neural networks. Although the stimulus reconstruction pattern can be easily transferred to the deep learning AAD methods \cite{de2020machine}, most of the works choose a more direct and end-to-end way, i.e. classify the speeches directly \cite{ciccarelli2019comparison,vandecappelle2021eeg,cai2021eeg,kuruvila2021extracting, su2022stanet}. For example, in \cite{cai2021eeg} and \cite{su2022stanet}, the authors build different attention mechanisms and apply them to the channel, band, or temporal of EEG to extract effective representation for AAD. However, these methods do not accumulate the prior knowledge of different views in the extraction of representation. As we mentioned before, the task-related information is contained in the attended speech and EEG, and ignoring such prior knowledge will hinder the performance of AAD methods. Different from those methods, we propose a multi-view auditory decoding method based on multi-view VAE, and use the TMC learning to accumulate the prior knowledge in the fusion of representation and learn an approximate task-related representation.

\subsection{Multi-View VAEs}
Recently, there has been a research interest in using VAE for self-supervised multi-view generative models, and produced a lot of important research progress \cite{kurle2019multi,wu2018multimodal,shi2019variational}. The greatest advantage of multi-view VAEs is that they can infer the complete representation given incomplete views of data. And the fundamental difference between these works is in the formulation of constructing the complete representation space, i.e. the complete posteriors. In MVAE \cite{wu2018multimodal}, the researchers use a product of single-view posteriors (Product-of-Experts, PoE \cite{hinton2002training}) to formulate the complete posterior. While in MMVAE \cite{shi2019variational}, the complete posterior is formulated using a mixture of single-view posteriors (Mixture-of-Experts, MoE). After that, several works have been proposed to improve the performance of MVAE and MMVAE \cite{wu2019multimodal,kurle2019multi,sutter2020multimodal,shi2020relating,daunhawer2021self}. In order to effectively combine the advantages of MVAE and MMVAE, MoPoE-VAE \cite{sutter2021generalized} use the Mixture-of-Products-of-Experts (MoPoE) which first conducts PoE on subsets of complete views and then form the complete posterior using MoE on these subsets.

\section{Methodology}
The primary purpose of AAD is to find the attended speaker in multiple speakers. We design our method based on the prior knowledge that the information about the attended speaker is contained in EEG and the attended speech. In contrast, unattended speech is unrelated to the goal when decoding auditory attention. Based on this idea, the main challenges are: 1) The method we use to construct the representation space; 2) How does our method reduce the interference of unrelated information while retaining the task-related information during the training? We will specify our method to address these two challenges in the following subsections.

\subsection{Decoding Auditory Attention with Multi-View VAE}
We construct our representation space using multi-view VAE. Specifically, we consider the EEG and speech stimuli (both the attended and unattended ones) as different views of data that may contain information about the subject's auditory attention, and use multi-view VAE to fuse different views into a common representation space. The overview of our method architecture is illustrated in figure \ref{fig:network}.

Given the raw EEG and speech stimuli, we extract different features from EEG and speech stimuli in the preprocessing stage. We extract the speech spectrogram using the short-time Fourier transform (STFT) from the lowpass-filtered raw speech stimuli. While for the EEG signal, we extract different frequency bands to construct a more comprehensive feature. We consider five EEG bands including the $\delta (1-4 \text{Hz}), \theta (4-8 \text{Hz}), \alpha (8-12 \text{Hz}), \beta (12-30 \text{Hz}), \text{and low } \gamma (30-50 \text{Hz})$ \cite{buzsaki2004neuronal}. The detailed implementation of the data preprocessing can be found in Section \ref{sec:exp}.

After data preprocessing, we mapped the different views of data into different single-view posteriors. All the single-view posteriors will be fused into a complete-view posterior $q_{\phi}(z|e, s_1, s_2)$ using Mixture-of-Products-of-Experts(MoPoE) \cite{sutter2021generalized}. Specifically, let $q_{\phi}(z|e)$ denote the posterior given EEG, $q_{\phi}(z|s_1)$ and $q_{\phi}(z|s_2)$ denote the posterior given different speeches, the MoPoE compute the complete posterior $q_{\phi}(z|e, s_1, s_2)$ as:
\begin{align}
    &q_{\phi}(z|\mathbf{X}_{k}) = p_{\theta}(z)\prod\limits_{m=1}^{N_{k}}q_{\phi}(z|x_{m}),\quad x_{m}\in\mathbf{X}_{k},\\
    &q_{\phi}(z|e, s_1, s_2) = \frac{1}{K} \sum\limits_{k=1}^{K} q_{\phi}(z|\mathbf{X}_{k}),
\end{align}
where $\mathbf{X}_{k}$ is a subset of complete view $\{e, s_1, s_2\}$ which has $N_k$ element, and $p_{\theta}(z)\sim \mathcal{N}(0, I)$ is an isotropic Gaussian. 

\begin{figure}[t]
    \centering
    \includegraphics[width=\linewidth]{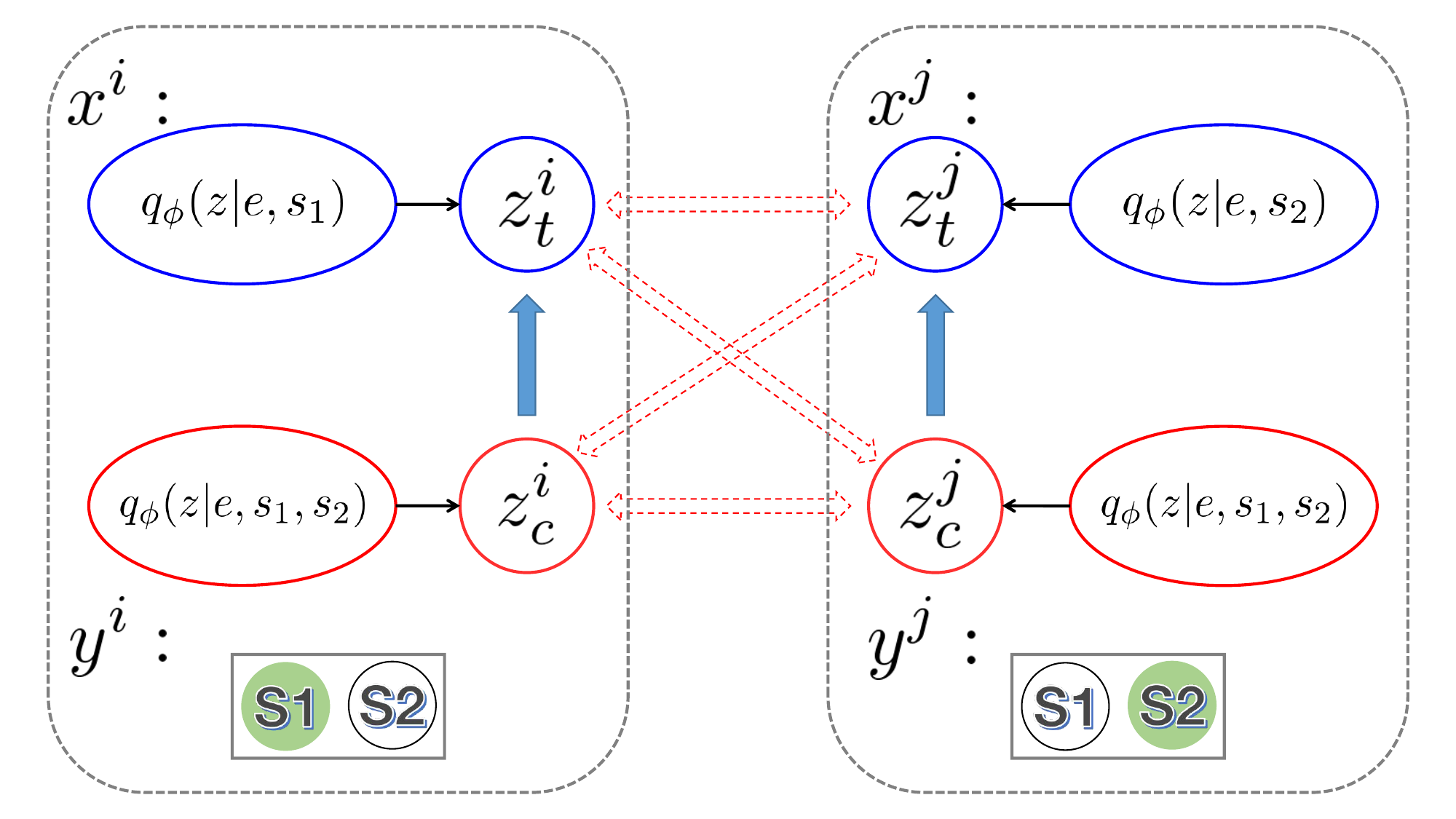}
    \caption{An example of TMC learning between two multi-view samples $x^i$ and $x^j$. We set the positive pair as the complete representation and the task-related representation in the same sample, and set the negative pair between different samples.}
    \label{fig:contrastive}
\end{figure}

The samples from the complete posterior will be fed into three decoders to reconstruct the original input, and the multi-view VAE will be trained by maximizing the evidence lower bound (ELBO):
\begin{align}
    &L_{ELBO} = \mathbf{E}_{z\sim q_{\phi}(z|e, s_1, s_2)}[\log (p_{\theta}(e, s_1, s_2|z))] \notag\\
    &\qquad\qquad - D_{KL}(q_{\phi}(z|e, s_1, s_2)\|p_{\theta}(z)),
\end{align}
where the $D_{KL}(\cdot)$ is the KL-divergence that is used to measure the statistical distance between complete posterior and the isotropic Gaussian $p_{\theta}(z)$.

The multi-view VAE can extract powerful representation in a self-supervision way. To approach the auditory attention in the representation space constructed by multi-view VAE, we apply a simple classifier in the complete representation space to separate samples with different attended speeches. The classifier contains a 3-layer MLP, which can map the samples from the complete posterior into a one-hot vector that indicates the attended speech in the input speeches.

We minimized the binary cross entropy (BCE) loss for the classifier during the training. Let the $\mathcal{C}(z)$ denote the classification result, and y is the label related to it, we compute classification loss as:
\begin{align}
    L_{\mathcal{C}} =& -\mathbf{E}_{z\sim q_{\phi}(z|e, s_1, s_2)} \left[y\log (\mathcal{C}(z)) \notag\right.\\
    &\left.+ (1-y)\log (1-\mathcal{C}(z))\right].
\end{align}

\subsection{Approximate the Task-Related Representation Using TMC}
Although the multi-view VAE can extract powerful representation from the complete view data, it has an inherent drawback in representation fusion. In fact, as we mentioned before, even if the unattended speech has task-unrelated information, we have no choice but to include it in the complete representation. Since removing the task-unrelated  unattended view needs the information from the label that is not available in the testing stage, it is impossible to get the task-related representations (posterior given EEG and attended speech) without ground-truth. 

To solve this problem and make better use of prior knowledge, we propose TMC learning that encourages multi-view VAE to learn an approximate task-related representation. A simple TMC instantiated between two multi-view samples $x^i$ and $x^j$ is shown in Figure \ref{fig:contrastive}.

\subsubsection{Task-Related Multi-View Contrastive (TMC) Learning.} In general, we use $\{x^{i}_{m}\}_{m=1}^{M} = \mathbf{X}^i$ to denote a sample from the general multi-view data which have M views. Moreover, we assume that there is a subset of views $\{x^{i}_{u}\}_{u \in U}=\mathbf{X}^i_{U} \subset \mathbf{X}^i$ that is task-unrelated. We refer to the single-view representations $z^{i}_{m} = f(x^{i}_{m})$ for the presentations extracted from the single-view data, and using $z^{i}_{c} = f(\mathbf{X}^i)$ for the complete representation which extracts from all the given views. Also, with the information from the gound-truth of the task, we have the task-related representation $z^{i}_{t} = f(\mathbf{X}^i_{T})$ which is fused by the representations from several single-views $\{x^{i}_{t}\}_{t \in T}=\mathbf{X}^i_{T} \subseteq \mathbf{X}$ that are related to the task. 

TMC uses contrastive learning to align the complete representation to the task-related one. Specifically, we compute the similarity of positive pair as:
\begin{equation}
    S_p = s_{c, t}(i, i) = exp(sim(z_{c}^{i}, z_{t}^{i})/\tau),
\end{equation}
where we choose cosine for $sim(\cdot, \cdot)$ and $\tau$ is the temperature hyperparameter. 

And for the negative pairs, we set the similarity between two different samples as the negative one. We consider the similarity between the complete representation and the task-related one as:
\begin{equation}
    S_n = s_{c, c}(i, j) + s_{c, t}(i, j) + s_{t, t}(i, j), i\neq j.
\end{equation}

So the TMC loss has the following form:
\begin{equation}
    L_{\mathcal{TMC}} = -\mathbf{E}_{z}\left[\log\frac{S_p}{S_n}\right].
\end{equation}

\subsubsection{Approximation Task-Related Representation.} As we mentioned before, the ideology task-related representation is unavailable in the testing stage, but it can be approximated through the complete representation by using the TMC learning.

Specifically, in the AAD task, where we use multi-view VAE as the backbone network, the single-view representation $z^{i}_{m} = f(x^{i}_{m})$ are sampled from the single-view posteriors learned by encoders related to different views:
\begin{equation}
    z^{i}_{m} \sim q_{\phi}(z|x_{m}), x_{m} \in \{e, s_1, s_2\}.
\end{equation}

Take the $s_1$ as the attended speech for an example, the complete representation $z^{i}_{c} = f(\mathbf{X}^i)$ and task-related representation $z^{i}_{t} = f(\mathbf{X}^i_{T})$ are extracted by the multi-view VAE from different fused posteriors:
\begin{align}
    &z^{i}_{c} \sim q_{\phi}(z|e, s_1, s_2), \\
    &z^{i}_{t} \sim q_{\phi}(z|e, s_1).
\end{align}

And the TMC can encourage the multi-view VAE to approximate the task-related representation by aligning the complete representation to the task-related one, which is fused by attended speech and EEG. We implement that by using the TMC loss to joint training multi-view VAE. So the loss function we used for the AAD task in our method is:
\begin{equation}
    L = -L_{ELBO}  + \alpha L_{\mathcal{C}} + \beta L_{\mathcal{TMC}},
\end{equation}
the $\alpha$ and $\beta$ are the weights of classification loss and TMC loss.

Although we propose TMC learning for the AAD task, we must point out that TMC is a general learning method. And the intuitive idea behind TMC can be applied to any multi-view data which bothered by the task-unrelated views.

In the implementation, we take the advantage of MoPoE that MoE and PoE are special cases of MoPoE. Specifically, when we constraint all the subsets of complete view only have single-view $\mathbf{X}_{1} = \{e\}, \mathbf{X}_{2} = \{s_1\} \mathbf{X}_{3} = \{s_2\}$, we can have the MoE posterior:
\begin{equation}
    q_{\phi}(z|e, s_1, s_2) = \frac{1}{3} \sum\limits_{m=1}^{3} q_{\phi}(z|x_{m}), x_{m}\in \{ e, s_1, s_2\},
\end{equation}
and when we constraint the MoPoE to have only one subset which is the complete view itself, we can have the PoE posterior:
\begin{equation}
    q_{\phi}(z|e, s_1, s_2) = p_{\theta}(z)\prod\limits_{m=1}^{3}q_{\phi}(z|x_{m}), x_{m}\in\{e, s_1, s_2\}.
\end{equation}
In Section \ref{sec:exp}, we give a thorough evaluation of our TMC learning with different fusion methods of multi-view VAE.

\begin{figure*}[t]
    \centering
    \includegraphics[width=\linewidth]{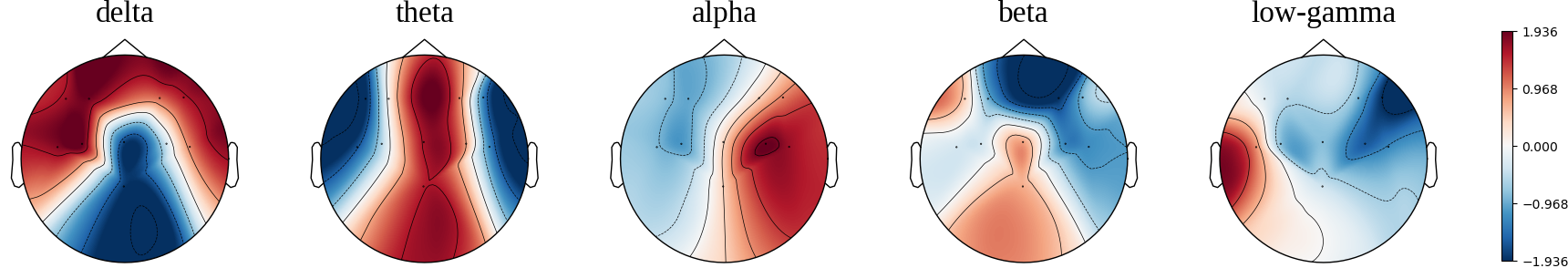}
    \caption{A topographic map of the EEG filter bank feature we used.}
    \label{fig:filter_bank}
\end{figure*}

\section{Experiments}\label{sec:exp}
\subsection{Experiment Setup}
\subsubsection{AAD Datasets.} We test our method on two popular AAD datasets. The first one is the {\bf KUL dataset} \cite{das2019auditory} which collects EEG data from 16 normal-hearing subjects in a soundproof and electromagnetically shielded room. The EEG data are collected by a 64-channel BioSemi ActiveTwo system at 8196 Hz sample rate. The stimuli are Dutch short stories narrated by different male speakers. To help the subjects focus on the experiment, the KUL dataset truncates the silences longer than 500 ms to 500 ms. The stimuli have two presentation conditions, HRTF (head-related transfer function) and dry. In the HRTF, stimuli applied to the subject's left and right ears are simulated by HRTF functions. While in the dry condition, the different story tracks are presented separately in the left ear or the right ear. We use the dry condition in our experiments, which has 4 trials with 6 mins duration for every subject. 

The {\bf DTU dataset} \cite{fuglsang2018eeg} contains EEG data from 18 subjects who take the experiment in a soundproof room. The EEG data are collected by a 64-channel BioSemi ActiveTwo system at 512 Hz sample rate. Different from the KUL dataset, the DTU datasets use Danish speeches narrated by a male and a female speaker. 
Every subject will experience 60 trials of speech stimuli, and every trial last for 50 s.

\subsubsection{Data Preprocessing.}
The speech stimuli are filtered and down-sampled before extracting the spectrogram. We first pass the speeches through a Chebyshev (type \uppercase\expandafter{\romannumeral2}) low-pass filter with 8 kHz cut-off frequency and downsampled the speeches to 16 kHz. Then we split the speeches into many decision windows and extract the spectrogram using the short-time Fourier transform with 32 ms Hann window and 12 ms hop length. 

For the EEG signal, we form a 3D filter bank by passing the EEG signal into the different Chebyshev (type \uppercase\expandafter{\romannumeral2}) band-pass filters, and concatenate the different frequency bands in one tensor. We use frequency bands of 1-4 Hz, 4-8 Hz, 8-12 Hz, 12-30 Hz, and 30-50 Hz which are known as $\delta$, $\theta$, $\alpha$, $\beta$, and low-$\gamma$ bands in EEG \cite{buzsaki2004neuronal}. While for the EEG channel, we follow the Joint CNN-LSTM \cite{kuruvila2021extracting} to use F7, F3, F4, F8, T7, C3, Cz, C4, T8, Pz instead of all the electrode. We refer to figure \ref{fig:filter_bank} for the topographic map of the EEG filter bank features.

In our experiments, we use two different decision window settings, 2 s and 3 s. With the longer decision window, the signal will contain more auditory attention information. Since our method is based on deep learning, we adopt the data augmentation by adding overlap between two windows. The overlap is set to 1 s for the 2 s decision window and 2 s for the 3 s decision window. The data volume after performing data augmentation is listed in Table \ref{tab:Data_volume}. 
\begin{table}[H]
    \centering
    \caption{Data volume of two datasets after data augmentation.}
    \label{tab:Data_volume}
    \begin{tabular}{l|ccc|ccc}
        \Xhline{1pt}
        \multirow{2}*{Dataset} & \multicolumn{3}{c|}{2s} & \multicolumn{3}{c}{3s} \\
        \cline{2-7} 
         & train & val & test & train & val & test \\
        \hline
        KUL & 18719  &  3120  &  3120  &  18719  &  3120  &  3120 \\
        DTU & 38879 & 6480 & 6480 & 38879 & 6480 & 6480 \\
        \Xhline{1pt}
    \end{tabular}
\end{table}

Since we use the same hop length for different decision windows, the total amount of data under different decision window lengths is the same. This setting can eliminate the impact on the performance caused by training data volume in different decision window lengths, especially for deep learning methods where data is a critical factor.

\subsubsection{Network Settings.} Our method is implemented based on Pytorch \cite{paszke2019pytorch}. For the encoder, we adapt the CNN part from Joint CNN-LSTM \cite{kuruvila2021extracting}, which uses 4 convolution layers for the EEG encoder and 5 for the speech encoder. We add one common linear layer and two private linear layers after the CNNs for the mean and variance of single-view posteriors. While for the decoder, we use a linear layer and several deconvolution layers (the same number as the single-view encoder) to reverse the process of encoding. For the classification part, we use a 3 layers MLP as our classifier.

We keep the network architecture identical in different decision windows. When decoding the auditory attention in 2 s decision windows, we just repeat and truncate the 2 s signal to make the input of the encoders have a 3 s length.

\subsubsection{Parameter Settings.} For all the experiments, we use $\alpha=1$, $\beta=1$ for the weight of classification loss and TMC loss, and set the temperature hyperparameter $\tau=1.5$. Moreover, for the dimension of representation learned by multi-view VAE, we use 128-dim for all the posterior fusion methods and conducted the experiment on a batch of 128 samples.


\subsection{Quantitative and Qualitative Results.}
In this part, we first evaluate the performance of task-related representation in the testing stage. Then we compare our method to several previous works in different decision windows on both datasets. We also include the MoPoE-VAE \cite{sutter2021generalized} and make a comparison of the representation similarity  to evaluate the effectiveness of our TMC learning. After that, we take close scrutiny to our TMC learning by evaluating TMC learning with different fusion methods. In all the tables except Table \ref{tab:ablation}, *
denotes the TMC-VAE performance is significantly better than the compared method (one-tailed unpaired
t-test, p<0.05).

\begin{figure}[H]
    \centering
    \subfigure[KUL 3s.]{
    \includegraphics[width=0.45\linewidth]{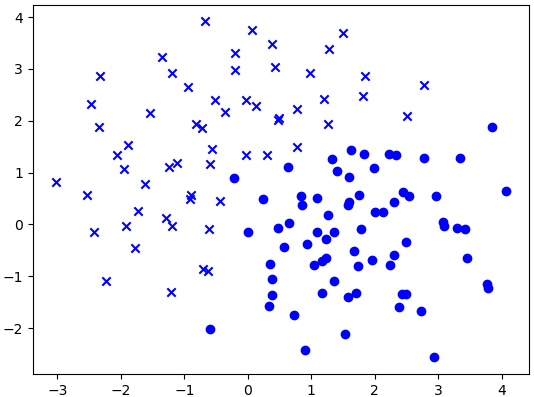}
    }
    \subfigure[KUL 2s.]{
    \includegraphics[width=0.45\linewidth]{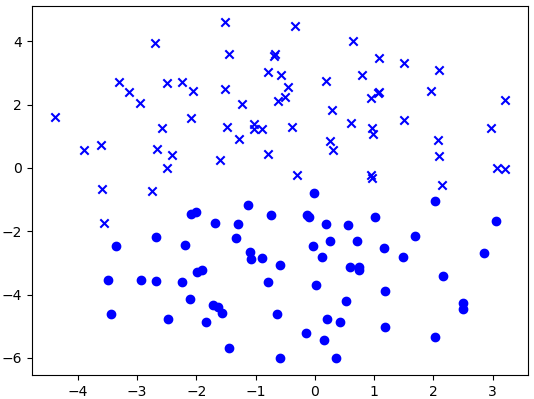}
    }
    \subfigure[DTU 3s.]{
    \includegraphics[width=0.45\linewidth]{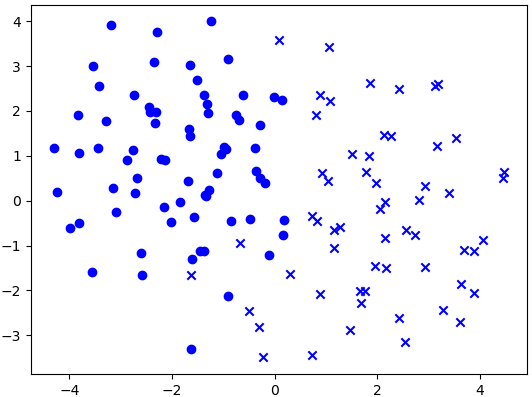}
    }
    \subfigure[DTU 2s.]{
    \includegraphics[width=0.45\linewidth]{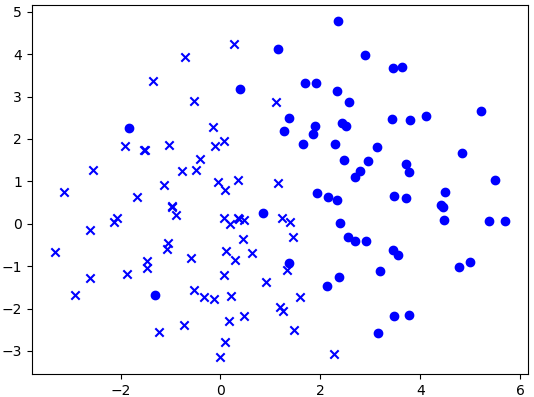}
    }
    \caption{Representation visualization of task-related representation. The $\cdot$ and $\times$ in the figure denote samples with different attended speeches.}
    \label{fig:target_representation}
\end{figure}

\subsubsection{Evaluation of Task-Related Representation.} To verify the reliability of our ideas, we first evaluate the decoding performance using task-related representation. We evaluate the performance in two aspects: 1) the decoding accuracy and 2) the visualization of task-related representation.

We use the label to construct task-related representation and classify the auditory attention based on it. The decoding performances are shown in Table \ref{tab:target_representation}. We find that the task-related representation can yield $100\%$ accuracy on both testing sets. We also visualize the task-related representation in the testing stage. As is shown in Figure \ref{fig:target_representation}, samples with different attended speeches are well separated in the task-related representation space.

Even though these results can not prove the performance of our method, the superior separation in the task-related representation space supports our motivation, which aims to construct an approximate task-related representation.

\begin{table}[H]
    \centering
    \caption{Decoding the auditory attention using task-related representation. } \label{tab:target_representation}
    \begin{tabular}{lll}
        \toprule
        Dataset  & 2s & 3s \\
        \midrule
        KUL &  100\%  & 100\%  \\
        DTU  & 100\% & 100\% \\
        \bottomrule
    \end{tabular}
\end{table}

\begin{figure*}[t]
    \centering
    \subfigure[MVAE.]{
    \includegraphics[width=0.3\linewidth]{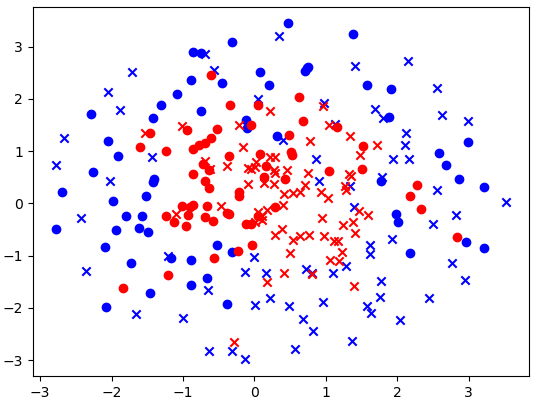}
    }
    \subfigure[MMVAE.]{
    \includegraphics[width=0.3\linewidth]{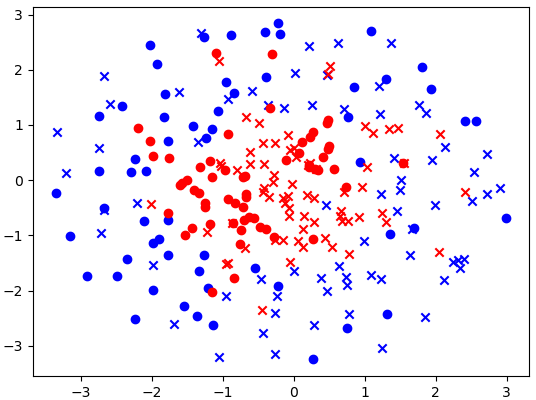}
    }
    \subfigure[MoPoE-VAE.]{
    \includegraphics[width=0.3\linewidth]{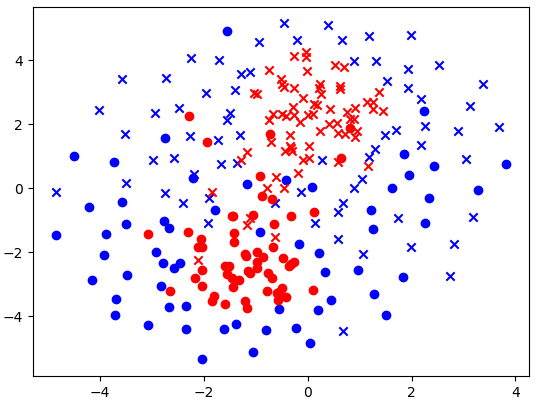}
    }
    \subfigure[MVAE+TMC.]{
    \includegraphics[width=0.3\linewidth]{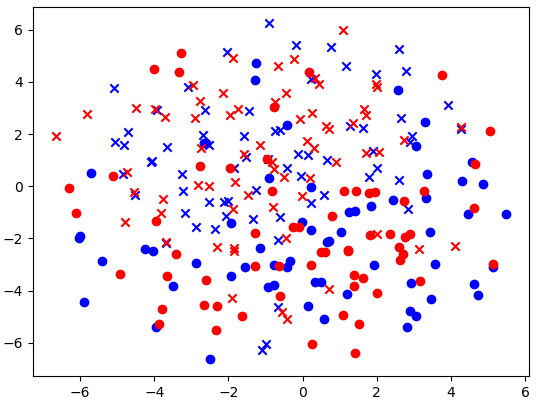}
    }
    \subfigure[MMVAE+TMC.]{
    \includegraphics[width=0.3\linewidth]{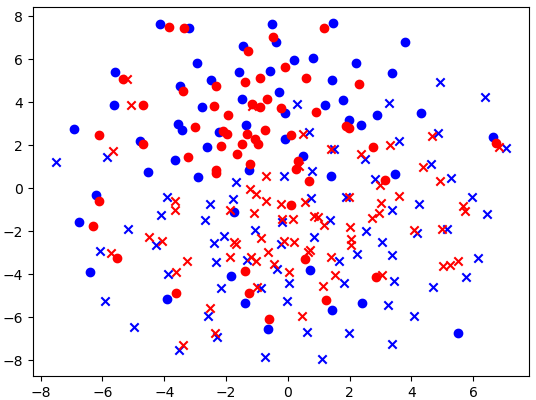}
    }
    \subfigure[TMC-VAE.]{
    \includegraphics[width=0.3\linewidth]{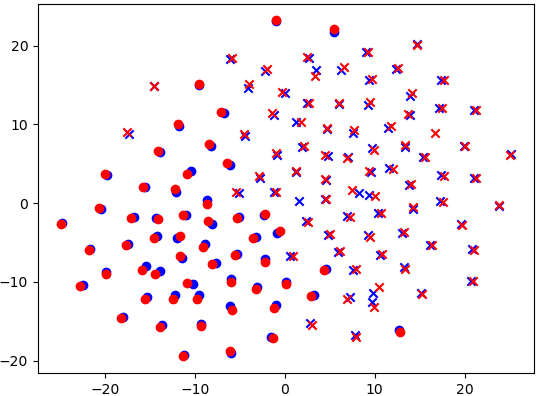}
    }
    \caption{Representation visualization of different multi-view VAE with or without TMC learning. We perform all the visualization on the KUL dataset with a 3 s decision window. In the first row, we present the visualization of representation without TMC learning. We use blue for the task-related representations and red for the complete ones. The $\cdot$ and $\times$ in the figure denote samples with different attended speeches.}
    \label{fig:TSNE}
\end{figure*}

\subsubsection{Comparative Study} In this part, we compare our method to several existing AAD methods. The joint CNN-LSTM \cite{kuruvila2021extracting} uses a convolution network (CNN) collaborative with a long short-term memory (LSTM) \cite{hochreiter1997long} to identify the attended speech in two input speeches. Deep CCA \cite{katthi2020deep} performs the correlation base AAD using the deep neural network with regularization. CNN-FC \cite{cai2021eeg} learn a discriminative representation for AAD by using the attention mechanism in their network, which is the state-of-the-art method on the KUL and the DTU datasets according to our knowledge. And the {\bf MoPoE-VAE} refers to using the same settings as our method but without TMC loss. We choose MoPoE as the fusion method of our approach in this study and use {\bf TMC-VAE} to represent this configuration.
\begin{table}[H]
    \centering
    \caption{Accuracy on the KUL dataset in different decision windows.}
    \label{tab:Acc_KUL}
    \begin{tabular}{lll}
        \toprule
        Method  & 2s & 3s \\
        \midrule
        Joint CNN-LSTM &  $72.7\%_{*}$  & $78.2\%_{*}$  \\
        Deep CCA & $66.4\%_{*}$ & $68.1\%_{*}$  \\
        CNN-FC  & \bf{86.9\%} & $89.7\%_{*}$ \\
        MoPoE-VAE & $83.6\%_{*}$ & $\underline{95.6\%}_{*}$ \\
        TMC-VAE & \underline{85.5\%} & \bf{96.6\%} \\
        \bottomrule
    \end{tabular}
\end{table}

We report the results in 2 s and 3 s decision windows on the KUL dataset in table \ref{tab:Acc_KUL}. Our TMC-VAE yields state-of-the-art result on the KUL dataset under 3 s decision window. Moreover, our TMC-VAE outperforms the existing methods by a large margin. Compared with the joint CNN-LSTM \cite{kuruvila2021extracting}, which has the same CNN encoder and more robust sequence data modeling capability with its LSTM module \cite{hochreiter1997long}, TMC-VAE can improve the decoding performance by $18.4\%$. Also, the comparison between MoPoE-VAE and existing methods can prove the advantages of using multi-view VAE to decode auditory attention. We notice that our method has a performance drop under a smaller decision window, but we must point out that our main contribution is using multi-view VAE and TMC learning to learn an approximate task-related representation, rather than carefully designing the networks. And our method can be easily adopted with a well-designed VAE backbone. 

We present the results on the DTU dataset in table \ref{tab:Acc_DTU}. The TMC-VAE also outperforms existing methods with a large margin under 3 s decision window. Although we use ordinary CNN architecture in our encoders and decoders, our method can perform comparable results with elaborately designed attention-based architecture (CNN-FC \cite{cai2021eeg}) under 2 s decision window. Also, in both datasets and all the decision windows, the comparison between the MoPoE-VAE and our method can demonstrate the effectiveness of TMC learning.

\begin{table}[H]
    \centering
    \caption{Accuracy on the DTU dataset in different decision windows.}
    \label{tab:Acc_DTU}
    \begin{tabular}{lll}
        \toprule
        Method  & 2s & 3s \\
        \midrule
        Joint CNN-LSTM &  $54.1\%_{*}$  & $55.6\%_{*}$ \\
        Deep CCA & $58.9\%_{*}$  & $60.7\%_{*}$ \\
        CNN-FC  & \bf{82.9\%} & $86.4\%_{*}$ \\
        MoPoE-VAE & $78.4\%_{*}$ & $\underline{91.5\%}_{*}$ \\
        TMC-VAE & \underline{80.8\%} & \bf{92.1\%} \\
        \bottomrule
    \end{tabular}
\end{table}

\subsubsection{Representation Similarity.} To make a better evaluation of the TMC learning and validate the performance improvement of TMC-VAE originates from the approximate representation, we compare the similarity between the task-related representation and the approximate one (complete representation) using the cosine similarity. We use the representations from two models in the testing stage in this part: 1) the MoPoE-VAE which trained without the TMC learning and 2) the TMC-VAE which trained with the TMC learning. The results are shown in Table \ref{tab:representation_sim} in which we can find that the similarity between the complete representation and task-related representation is increased significantly.
\begin{table}[H]
    \centering
    \caption{Representation similarity.}
    \label{tab:representation_sim}
    \begin{tabular}{lllll}
        \toprule
        Method  & KUL 2s & KUL 3s & DTU 2s & DTU 3s\\
        \midrule
        MoPoE-VAE & $0.037_{*}$ & $0.041_{*}$ & $0.008_{*}$ & $0.008_{*}$ \\
        TMC-VAE & {\bf 0.356} & {\bf 0.361} & {\bf 0.311} & {\bf 0.291} \\
        \bottomrule
    \end{tabular}
\end{table}

\subsubsection{Effectiveness of TMC with Different Fusion Methods.}\label{sec:effect-TMC}
To make a thorough study of the effectiveness of our TMC learning, we present the ablation study of TMC learning with different multi-view VAEs. We choose three typical multi-view VAEs here: 1) MVAE \cite{wu2018multimodal} which uses PoE to fuse the single-view posteriors 2) MMVAE \cite{shi2019variational} which proposes MoE in the fusion of complete posteriors and 3) MoPoE-VAE \cite{sutter2021generalized} which take the advantages from PoE and MoE, and proposed a general fusion modal. 
\begin{table}[h]
    \centering
    \caption{AAD accuracy of different fusion methods on the KUL and the DTU datasets. * denotes the performance with TMC is significantly better than without TMC (one-tailed unpaired t-test,
p<0.05).}
    \begin{tabular}{lllll}
        \toprule
        \multirow{2}*{Method}  & \multicolumn{2}{c}{KUL} & \multicolumn{2}{c}{DTU} \\
        \cmidrule(lr){2-3}\cmidrule(lr){4-5}
        & 2s & 3s & 2s & 3s \\
        \midrule
        MVAE & $83.6\%_{*}$ & $93.9\%_{*}$ & $77.1\%_{*}$ & $89.9\%_{*}$ \\
        MVAE+TMC & 84.5\% & 94.3\% & 78.2\% & 90.7\% \\
        MMVAE & $84.4\%_{*}$ & $92.7\%_{*}$ & $78.3\%_{*}$ & $87.7\%_{*}$\\
        MMVAE+TMC & 84.7\% & 93.1\% & 78.8\% & 88.4\%\\
        MoPoE-VAE & $83.6\%_{*}$ & $95.6\%_{*}$ & $78.4\%_{*}$ & $91.5\%_{*}$\\
        TMC-VAE & {\bf 85.5\%} & {\bf 96.6\%} & {\bf 80.8\%} & {\bf 92.1\%}\\
        \bottomrule
    \end{tabular}
    \label{tab:ablation}
\end{table}

 It is shown in table \ref{tab:ablation} that TMC learning can encourage multi-view VAE to learn an approximate task-related representation with different posterior fusion formulas. Also, the TMC-VAE yields the best results, which also gives quantitative support for the advantage of choosing MoPoE in TMC-VAE.

\subsubsection{Representation Visualization.}
In this part of the study, we present some qualitative results in figure \ref{fig:TSNE} to give intuitive evidence of the effectiveness of TMC learning. Specifically, we visualize the representation learned by different multi-view VAEs with or without TMC learning. We use the same multi-view VAEs in Section \ref{sec:effect-TMC}.

All the visualizations are yields in the testing stage on the KUL dataset under a 3 s decision window. To perform the visualizations, we first feed a batch of data into the trained encoders to extract the 128-dim representations, and then map these representations to 2-dim using t-SNE \cite{van2008visualizing}. We apply blue for the task-related representations and red for the complete ones, and distinguish the samples with different attended speeches using circle and cross. 

As shown in figure \ref{fig:TSNE}, the task-related representations (in blue) are more separable in the representation space than the complete one (in red). While TMC learning can encourage the multi-view VAEs to learn a more separable complete representation by aligning it to the task-related one. We also observed that the TMC-VAE has the most prominent consistency and separability in all the visualizations, which suggests that MoPoE is the optimal choice of the fusion formula in our method. We report the decoding performance of different posterior fusion methods in the next part study.

\section{Conclusions.}
In this work, we first introduce a multi-view VAE and a classifier to learn the multi-view representation for AAD efficiently. Then, inspired by Broadbent's filter model, we define the task-related representation in the AAD task, and propose the TMC learning to encourage the complete representation aligning with the task-related one. Finally, the experiments on the KUL and the DTU datasets prove the advantages of our method. 

\begin{acks}
This work was supported in part by the Scientific and Technological Innovation (STI) 2030–Major Projects under Grant 2021ZD0201503; in part by the National Natural Science Foundation of China under Grant 82272072; and in part by the CAAI-Huawei MindSpore Open Fund.
\end{acks}

\bibliographystyle{ACM-Reference-Format}
\balance
\bibliography{TMC}
\end{document}